\documentclass[]{spie}

\title{Experimental validation of photonic lantern imaging and wavefront sensing performance}

\author[a]{Aditya R. Sengupta}
\author[b]{Vincent Chambouleyron}
\author[a]{Jordan Diaz}
\author[a]{Matthew DeMartino}
\author[a]{Rebecca Jensen-Clem}
\author[c]{Benjamin L. Gerard}
\author[c]{Michael J. Messerly}
\author[c]{Paul Pax}
\author[a]{Daren Dillon}
\author[a]{Kevin Bundy}
\author[d]{Maria Cuevas}
\author[e]{Sylvain Cetre}
\author[a]{Bruce Macintosh}
\author[f]{Caleb Dobias}
\author[f]{Tara Crowe}
\author[f]{Stephen S. Eikenberry}
\author[f]{Rodrigo Amezcua-Correa}
\author[f]{Stephanos Yerolatsitis}

\affil[a]{Department of Astronomy \& Astrophysics, University of California, Santa Cruz, CA 95064, USA}
\affil[b]{Laboratoire d'Astrophysique de Marseille}
\affil[c]{Lawrence Livermore National Laboratory, Livermore, CA 94550, USA}
\affil[d]{Wakea Consulting, Durham, UK}
\affil[e]{Department of Astronomy, Columbia University, 550 W 120th Street, New York, NY 10027}
\affil[f]{CREOL, The College of Optics \& Photonics, University of Central Florida, Orlando, FL
32816, USA}

\authorinfo{Further author information: (Send correspondence to A.S.)\\A.S.: E-mail: adityars@ucsc.edu}

\usepackage{preamble}

\begin{document} 
\maketitle
\pagenumbering{gobble}
\begin{abstract}
Photonic lanterns (PLs) are fiber-based waveguides that are capable of focal-plane wavefront sensing while simultaneously directing  light to downstream science instruments. The optimal choice of wavefront reconstruction algorithm has yet to be determined, and likely depends on the particular observing scenario under consideration. Previous work in simulation suggests that PLs can be used for nonlinear wavefront sensing for several applications, including sensing the low-wind effect and correcting large-amplitude aberrations. We present the design of muirSEAL (miniature IR SEAL), a testbed designed to test PL wavefront reconstruction over Zernike modes and segmented-mirror offsets. We demonstrate throughput and linear wavefront reconstruction at multiple f-numbers. We further present initial laboratory imaging of a new photonic lantern fabricated at Lawrence Livermore National Laboratory.  

\end{abstract}


\section{Introduction}

Photonic lanterns (PLs) act as focal-plane wavefront sensors that can direct light downstream to science instruments, which is well suited to the needs of high-contrast imaging systems on future  ground- and space-based observatories due to their small form factor. The wavefront sensing performance of PLs has been assessed in simulation\cite{Lin22} and demonstrated in laboratory settings and on sky\cite{Norris20, Xin24, Sengupta2024, Lin25}. However, the full capabilities and limitations of PLs as wavefront sensors have yet to be explored in a laboratory setting. For example, PLs have an optimal f-number for injection based on their numerical aperture and the size of the input PSF relative to the multi-mode diameter at the input end of the PL. It is unknown whether this is also the optimal f-number for wavefront sensing; we expect a tradeoff between throughput and wavefront sensing performance, since wavefront sensing for higher spatial orders requires more light to be coupled from the further-out/dimmer Airy rings of a perfect PSF.

We previously assessed the AO performance of a PL built at CREOL at the University of Central Florida on the SEAL testbed at UC Santa Cruz \cite{Sengupta2024,SEAL}. However, this PL was designed for an input wavelength of 1.55 $\mu$m, whereas at the time SEAL was a single-wavelength testbed at $0.633 \mu$m. The outputs were often visibly multi-modal and it was unclear whether differences in performance from simulations were due to the wavelength mismatch or other aspects of the experiment.

We present the design of muirSEAL (the \textbf{m}iniat\textbf{u}re \textbf{i}nfra\textbf{r}ed Santa cruz Extreme AO Laboratory), a separate testbed in the same enclosure as the main SEAL bench \cite{SEAL,SEAL2} intended to test PLs with a design wavelength of 1.55 $\mu$m (see also Jensen-Clem \textit{et al.} in these proceedings). This work enables PL wavefront reconstruction of aberrations due to several different simulated sources at once (e.g. atmospheric aberrations, segment piston offsets, and the low wind effect), while varying both the reconstruction method and the f-number of the light injected at the PL entrance. 

We further present the initial testing of another photonic lantern made by the fiber-optics lab at Lawrence Livermore National Laboratory, consisting of seven PLs in one device. 

\section{Testbed setup}

\begin{figure}
    \centering
    \includegraphics[width=\textwidth]{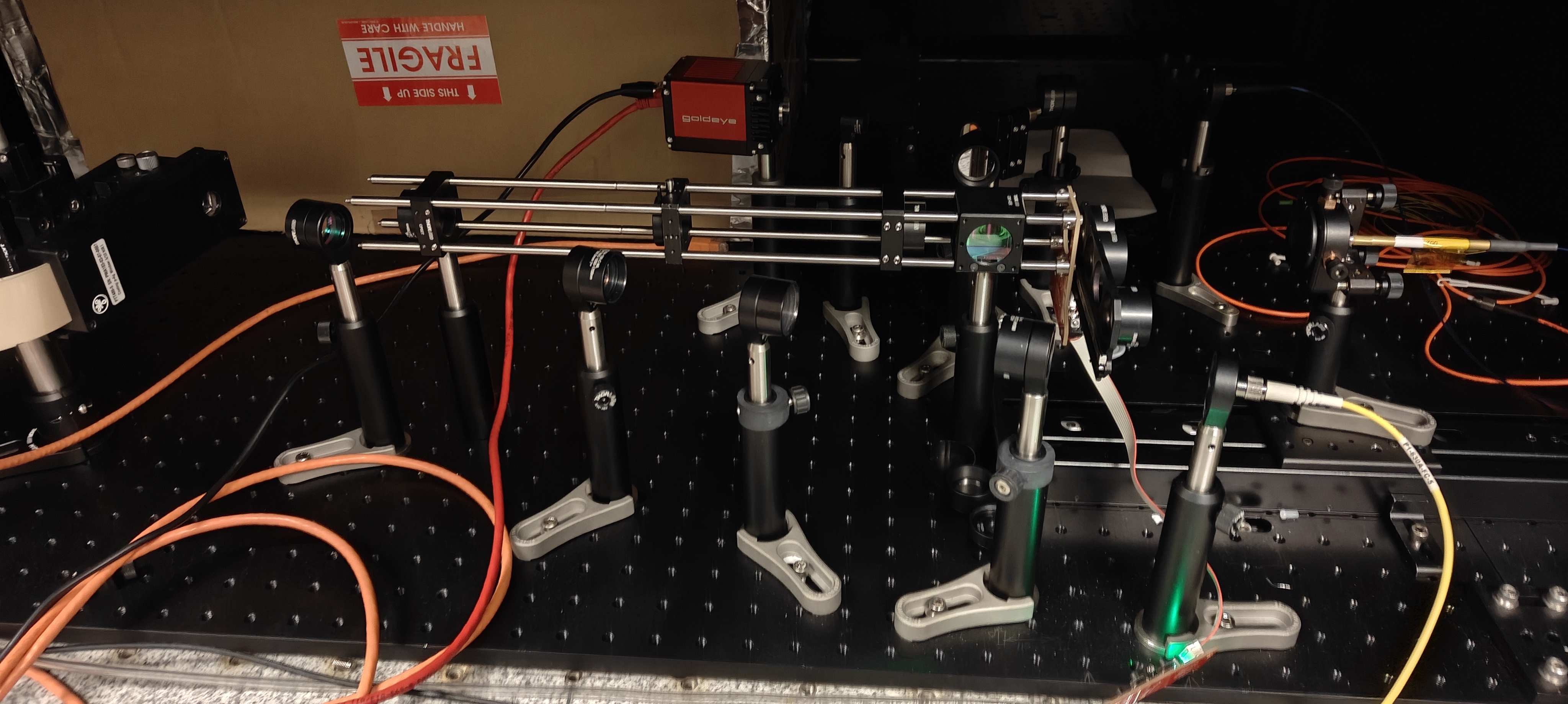}
    \caption{The muirSEAL testbed enables testing PL injection and wavefront reconstruction at several different PSF sizes.}
    \label{fig:muirseal_bench}
\end{figure}

\begin{figure}[!htbp]
    \centering
    \includegraphics[width=\textwidth]{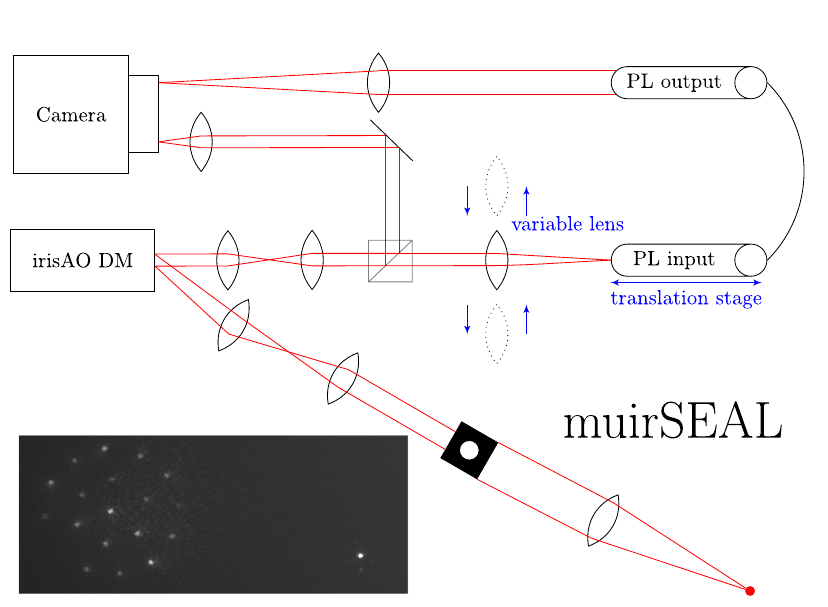}
    \caption{The muirSEAL testbed optical layout.}
    \label{fig:muirseal}
\end{figure}
\begin{figure}[!htbp]
    \centering
    \includegraphics[width=0.5\textwidth]{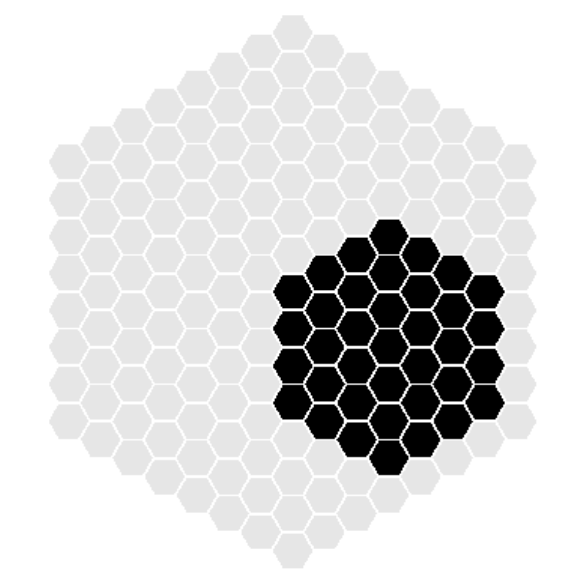}
    \caption{The subaperture on the segmented DM that is illuminated by muirSEAL.}
    \label{fig:in_beam_segments}
\end{figure}


muirSEAL uses a 1.55$\mu$m laser source (Thorlabs KLS1550). Input light goes through a 1cm diameter iris and is incident on a 169-actuator IrisAO deformable mirror (DM) that is used to simulate both atmospheric turbulence and primary-mirror segment phasing (Cuevas \textit{et al.} in these proceedings\cite{Cuevas25}). The beam is undersized relative to the DM, so it sees four rings of segments out of the total of eight. Figure~\ref{fig:in_beam_segments} shows which segments are illuminated. After the DM, light is split between a path reimaging the PSF, and a path that injects light into the PL. 

We use a Thorlabs Elliptec four-position slider to change the lens focusing light into the PL entrance, and a linear translation stage to repeatably move the PL's input end to the focus of each lens. This allows us to test injection at f-numbers between 2 and 12.

Light from both paths is reimaged onto our detector, which is a Goldeye CL-030 VSWIR TEC1.

\section{Testbed experiments}

\subsection{Imaging setup}

We keep the laser power fixed at 0.06 mW and set the exposure time per f-number in order to maximize counts without saturating the detector on the PL image at the flat position. The exposure times range from 0.1 ms for f/4 to 5 ms for f/12. To average out fluctuations in the laser output, we median-combine 20 frames on the detector per measurement.

\subsection{Injection efficiency testing}

\begin{figure}
    \centering
    \includegraphics[width=0.7\linewidth]{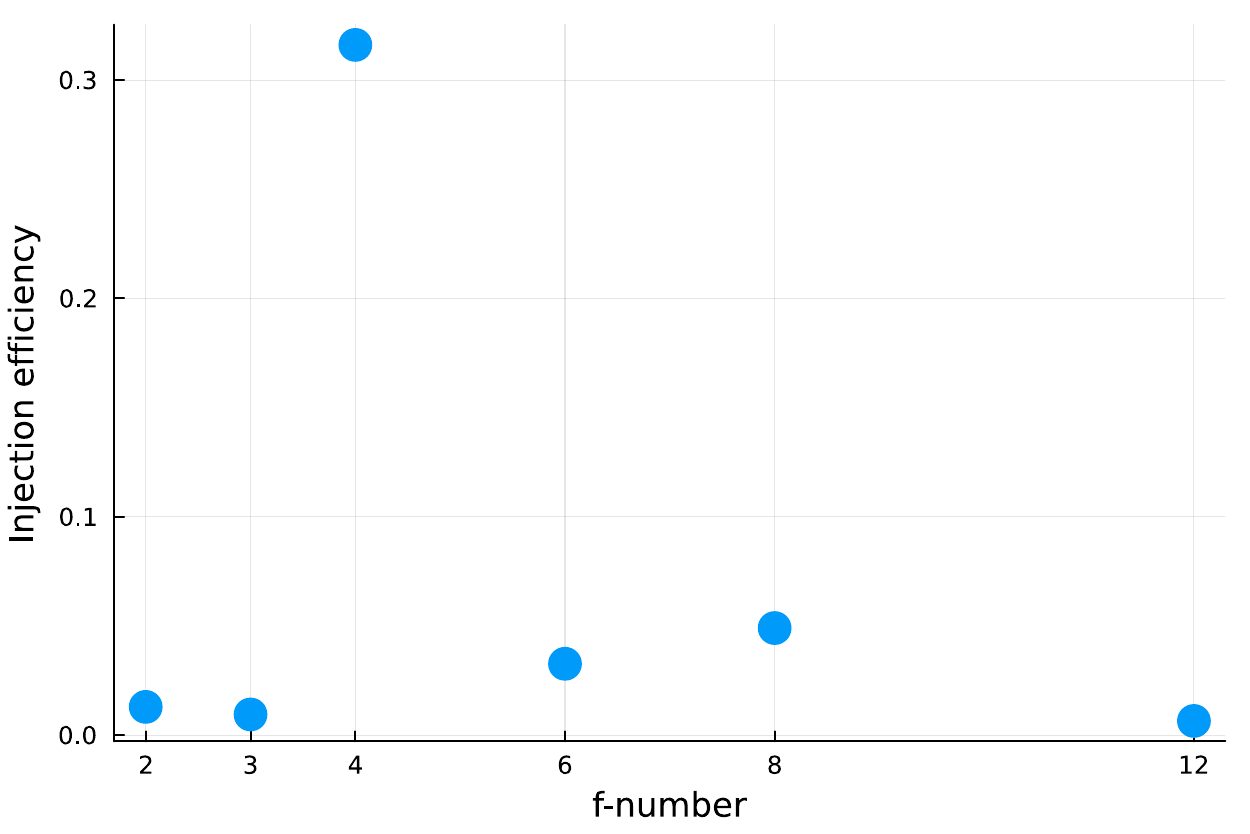}
    \caption{Injection efficiency as a function of f-number as measured on muirSEAL.}
    \label{fig:throughput}
\end{figure}

We measure the injection efficiency of the PL at six different f-numbers: f/2, 3, 4, 6, 8, and 12. Rather than comparing counts through the PL to those through a multi-mode fiber at the same location as in previous tests of PL throughput\cite{Lin2022coupling}, we measure injection efficiency by dividing the total counts through the PL at each f-number by the total counts at the input. We measure the latter by temporarily moving the detector to the position of the PL input, and we adjust tip/tilt and focus per f-number to maximize counts through the PL. At certain f-numbers the PL image is not visible at the exposure time used for the corresponding input; in these cases we scale up the exposure time by the smallest possible factor such that all the ports are visible by eye, and scale down the counts on this image by this exposure time scaling factor. 

Figure~\ref{fig:throughput} shows the resulting injection efficiency as a function of the f-number at the PL input. We note that PLs at a single wavelength in the near IR show coupling efficiencies of 97-98\% \cite{Moraitis2021}; however, our best injection efficiency is 31.6\%. This is likely to be characteristic of the injection efficiency of the pre-PL optics rather than the inherent properties of the lantern. The presence of the peak at f/4 is consistent with this PL's numerical aperture of 0.11 (corresponding to an f-number of $1 / (2 \times \text{NA}) = 4.54$. We expect a throughput of 76\% at f/4; from the design, we expect 79\% ($(4/4.5)^2$) efficiency, together with 97\% efficiency as previously characterized. This is a difference of around 2.5x. This is likely to be fixable with pre-PL alignment rather than being an inherent property of this lantern.

We note the double-peak structure (with peaks at f/4 and f/8) and the sharp dropoff around f/4. Both of these features are unexpected, since previous results show only a single peak and a less drastic reduction in throughput at higher f-numbers than the peak \cite{Lin2022coupling}. Given that the PL throughput is highly sensitive to tip/tilt variations, our results could be due to the fact that the x/y position of the input was set slightly differently for each f-number. We expect that refining the x/y input positioning procedure could change the results shown in Figure~\ref{fig:throughput}. Generating coupling maps for the PL at each f-number, as well as more detailed modeling of this particular design, would help to improve the consistency of the alignment.

\subsection{Linear reconstruction}

At each f-number, we make a modal interaction matrix using 37 Zernike modes (matching the number of DM segments imaged). Since we imaged an off-center subaperture in order to avoid two dead actuators, we build our own Zernike solution by expressing each Zernike phase screen in terms of the piston, tip, and tilt of each segment using \textit{hcipy}\cite{hcipy}. We confirm that the segments are well centered and that the Zernike solution is valid by observing the PSF shape while each mode is applied.

We extract signals from the PL using aperture photometry around each port. We visually identify approximate centers for each port in the detector, and refine these positions by taking a center of mass within an image cutout around the approximate center. We then make circular masks of a fixed radius (varying as a function of alignment, typically around 6 px or 30 $\mu$m) for each port. The PL spots are sharply defined by eye, making it easy to identify mask misalignments. Within each mask, we capture the port as well as a small annulus of dark pixels around it. Due to the small exposure time and low gain, the ports are easily distinguished from the surrounding un-illuminated pixels, and the additional noise may be neglected. We take as our measurement the sum of the counts through each port, i.e. the sum of the image multiplied by each mask in turn, and we normalize this (divide each measurement by the sum of all 19 measurements) to obtain a WFS signal.

We build a linear model of the PL at each f-number by applying positive and negative pokes in each Zernike mode and forming an interaction matrix out of the resulting differential port intensities on the PL. We invert this to form a command matrix, using a singular value threshold of 1/30. We then assess reconstruction quality by taking linear sweeps in each Zernike mode, and assessing the reconstruction in the injected mode as well as in others. We truncate this test at 9 modes for comparison with our earlier PL lab results\cite{Sengupta2024}.

\begin{figure}
    \centering
    \begin{subfigure}{0.325\linewidth}
        \centering
        \includegraphics[width=\linewidth]
        {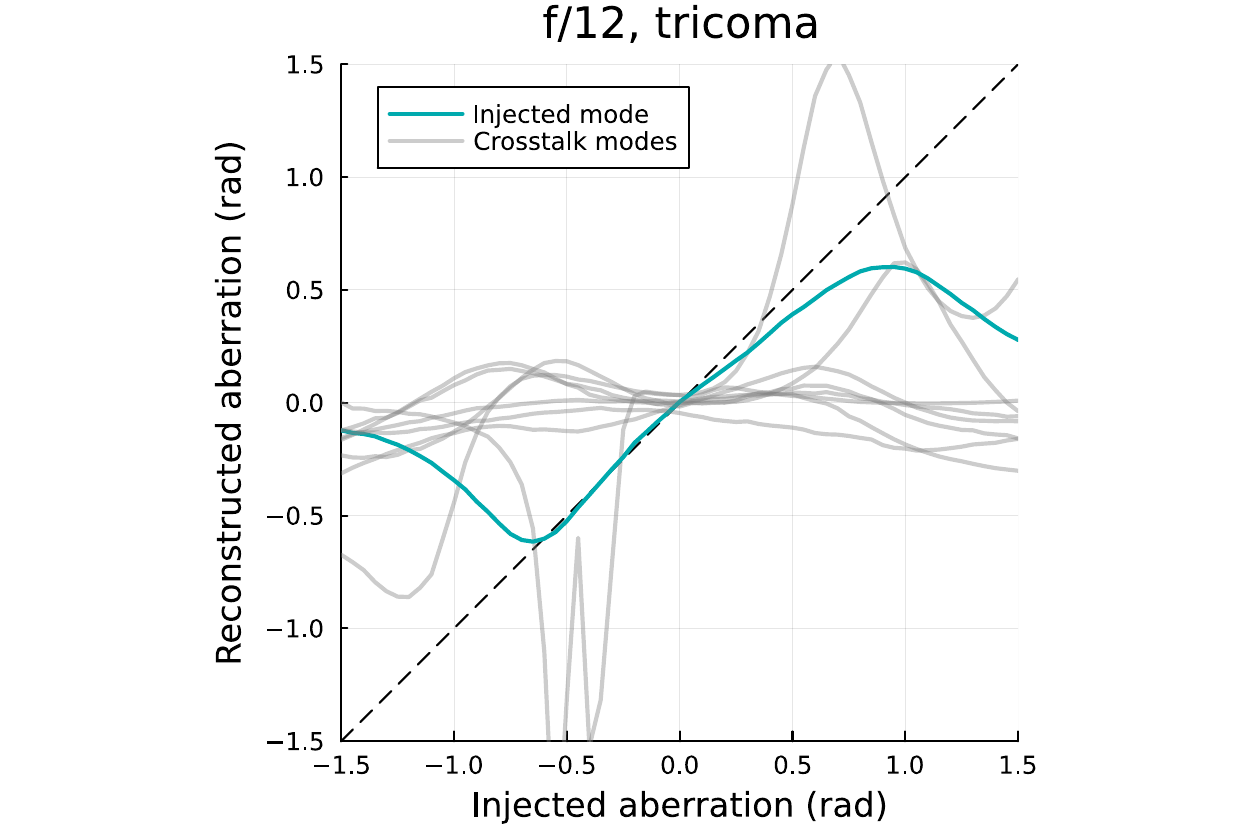}
        \caption{Representative linearity curve} \label{fig:representative_linearity_curve}
    \end{subfigure}\hspace*{\fill}
    \begin{subfigure}{0.33\linewidth}
        \centering
        \includegraphics[width=\linewidth]{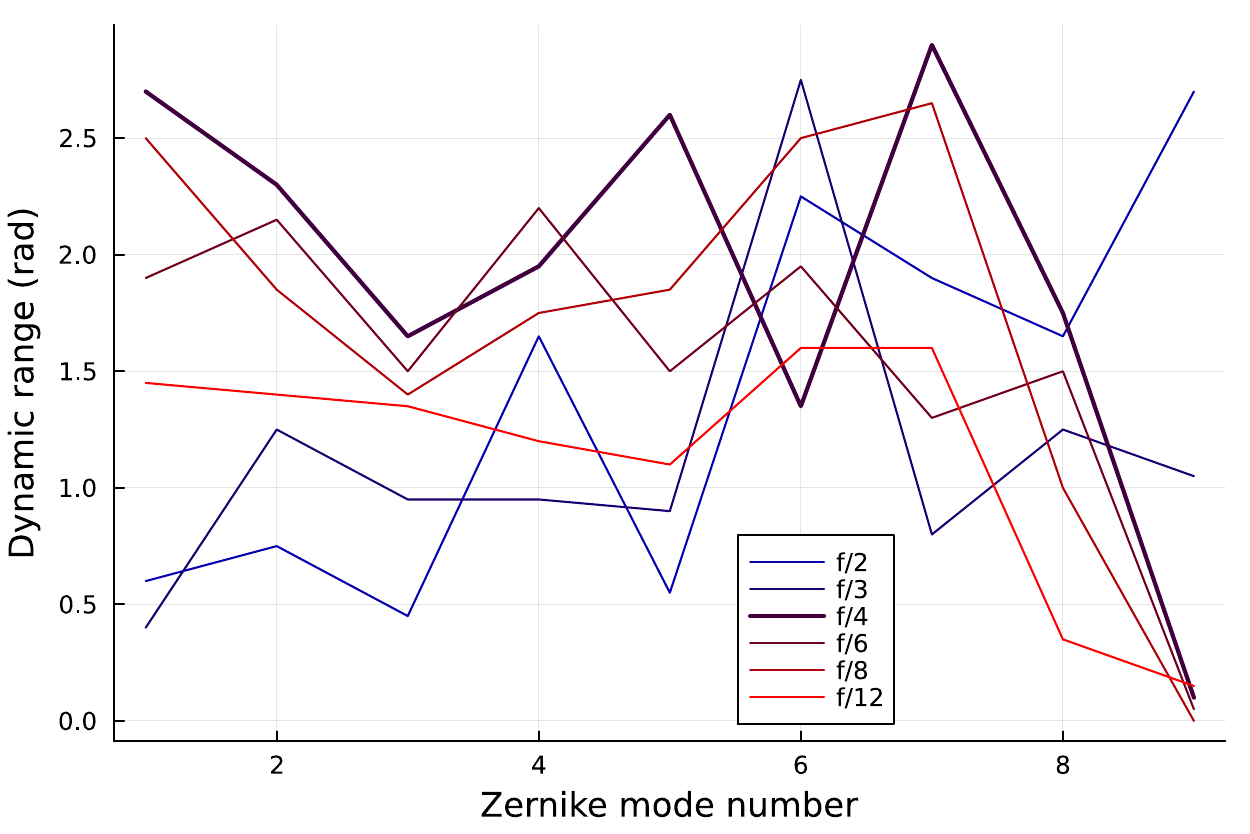}
        \caption{Dynamic ranges}
        \label{fig:dynamic_ranges}
    \end{subfigure}
    \begin{subfigure}{0.33\linewidth}
        \centering
        \includegraphics[width=\linewidth]{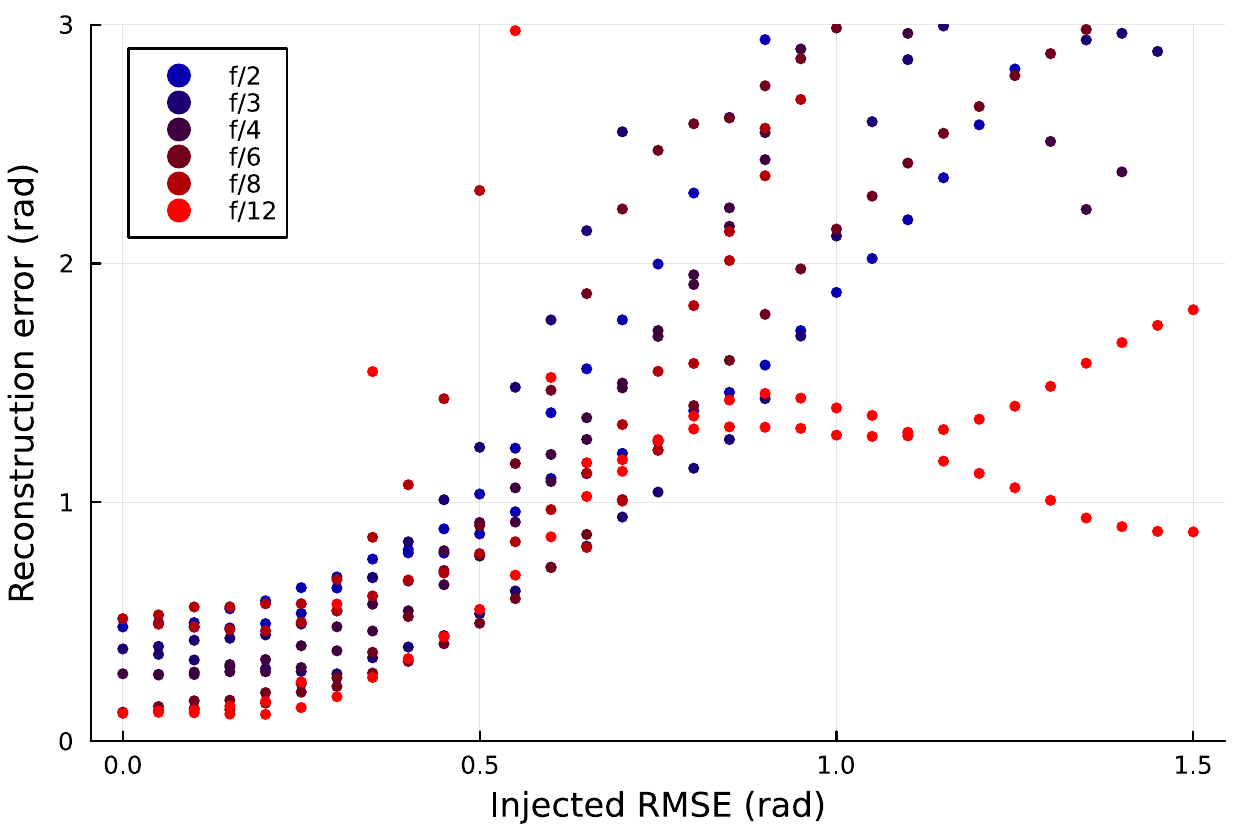}
        \caption{Total linear reconstruction errors}
        \label{fig:nonlinear_reconstruction}
    \end{subfigure}
    \caption{Linear reconstruction performance of the PL on muirSEAL.}
    \label{fig:linear_recon}
\end{figure}

Figure~\ref{fig:representative_linearity_curve} shows an example linearity curve for one f-number and one aberration. The high-opacity line is the injected mode and the low-opacity lines are crosstalk, i.e. the reconstructions in modes that were not applied. We observe wide linear ranges, but strong crosstalk except in a range close to zero. We identify a dynamic range by finding the first overturning in the injected mode in both directions. We define an overturning as the first reconstructed coefficient that is less than the previous ones when the injected aberration coefficient is strictly increasing; this represents the first point at which the PL is unable to distinguish between the applied aberration and a smaller one. 

Figure~\ref{fig:dynamic_ranges} shows the dynamic ranges for all considered f-numbers and all Zernike modes. f/4 is in bold to denote that the throughput at this position was much higher than the others. We identify no strong trends in dynamic ranges, other than a possible reduction in dynamic range for f/3 and increase in dynamic range for the coma and tricoma modes (6-9). This may be indicative of higher spatial frequency features on the PSF falling outside the PL input at these higher f-numbers.

Figure~\ref{fig:nonlinear_reconstruction} shows the overall reconstruction error at each f-number as a function of the input wavefront error. We note that the largest f-numbers (f/8 and f/12) perform better at large wavefront errors, but since these points lie well outside the dynamic range of the PL, this is not likely to be relevant for practical wavefront reconstruction. We observe no strong trends as a function of f-number for small input wavefront errors.

We note significantly larger reconstruction crosstalk than with previous work on the main SEAL testbed\cite{Sengupta2024}. Reconstruction crosstalk is seen when modes other than the one that was applied are present in the reconstructed wavefront, and are likely characteristic of alignment or of the choice of modal basis. The large amount of reconstruction crosstalk in this case may be due to the comparatively lower-order DM on muirSEAL versus the DM on SEAL, differences in the alignment between the six f-numbers, and/or differences in the lanterns used. Further testing of the linear wavefront reconstruction pipeline, as well as the development of nonlinear methods, will be beneficial to resolve this.

\section{The LLNL photonic lantern}

\begin{figure}
    \centering
    \includegraphics[width=\textwidth]{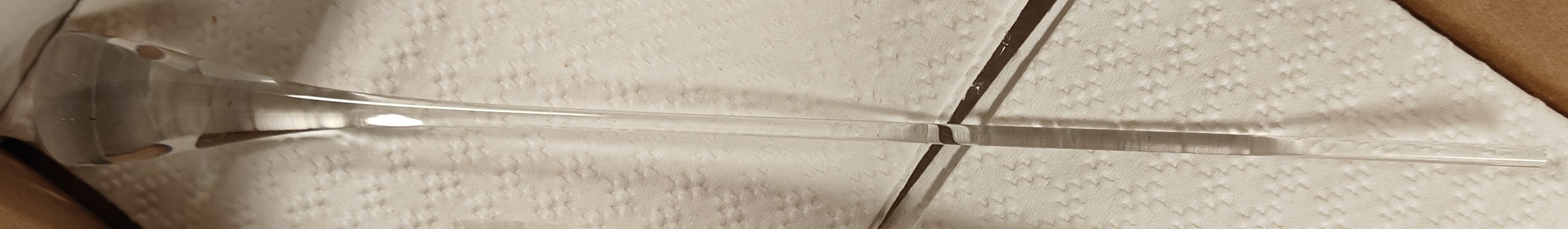}
    \caption{The LLNL photonic lantern}
    \label{fig:llnl_lantern}
\end{figure}

In the previous section, we explored the wavefront sensing capabilities of a single PL as a function of f-number. We now aim to explore how the wavefront sensing capabilities are affecting manufacturing differences between lanterns. We collaborated with the fiber-optics laboratory at LLNL to make a new type of optical waveguide. This device (shown in Figure~\ref{fig:llnl_lantern}) consists of seven 19-port PLs made to a design wavelength of 0.633 $\mu$m. Four of these lanterns are standard, meaning all output ports have the same diameter, and three are hybrid, meaning some output ports are differently sized.  More details about the fabrication and characterization of this lantern will be presented in a future paper. Beyond the usual features of a PL, this device enables simultaneous imaging of the same PSF onto multiple lanterns, allowing us to extract higher-order modal information; it also allows for testing of variation in PL performance as a result of manufacturing differences.

\begin{figure}
    \centering
    \begin{subfigure}{0.42\linewidth}
        \centering
        \includegraphics[width=\linewidth]{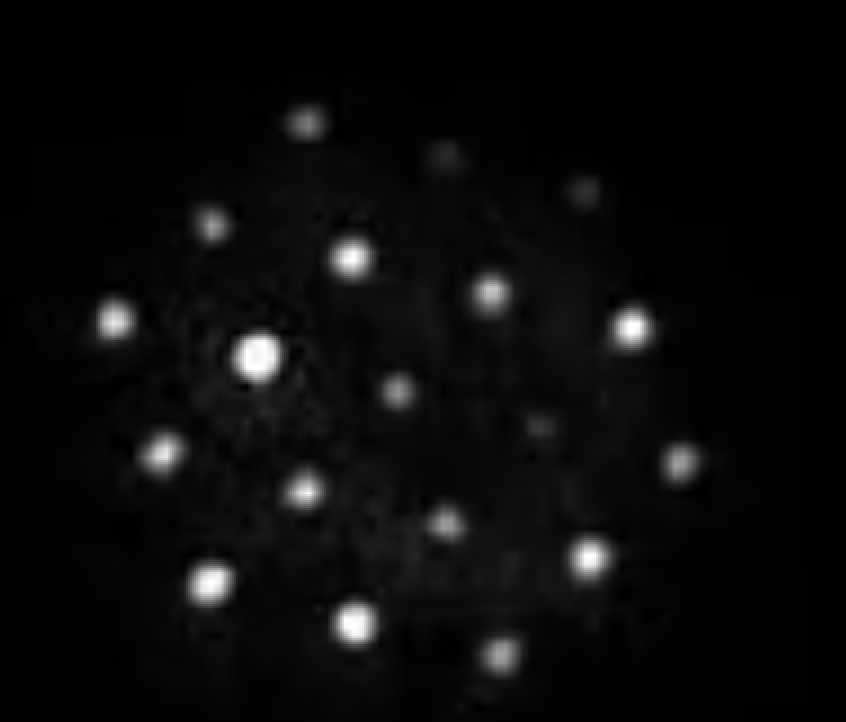}
        \caption{Laboratory image}
        \label{fig:real}
    \end{subfigure}\hspace*{\fill}
    \begin{subfigure}{0.53\linewidth}
        \centering
        \includegraphics[width=\linewidth]
        {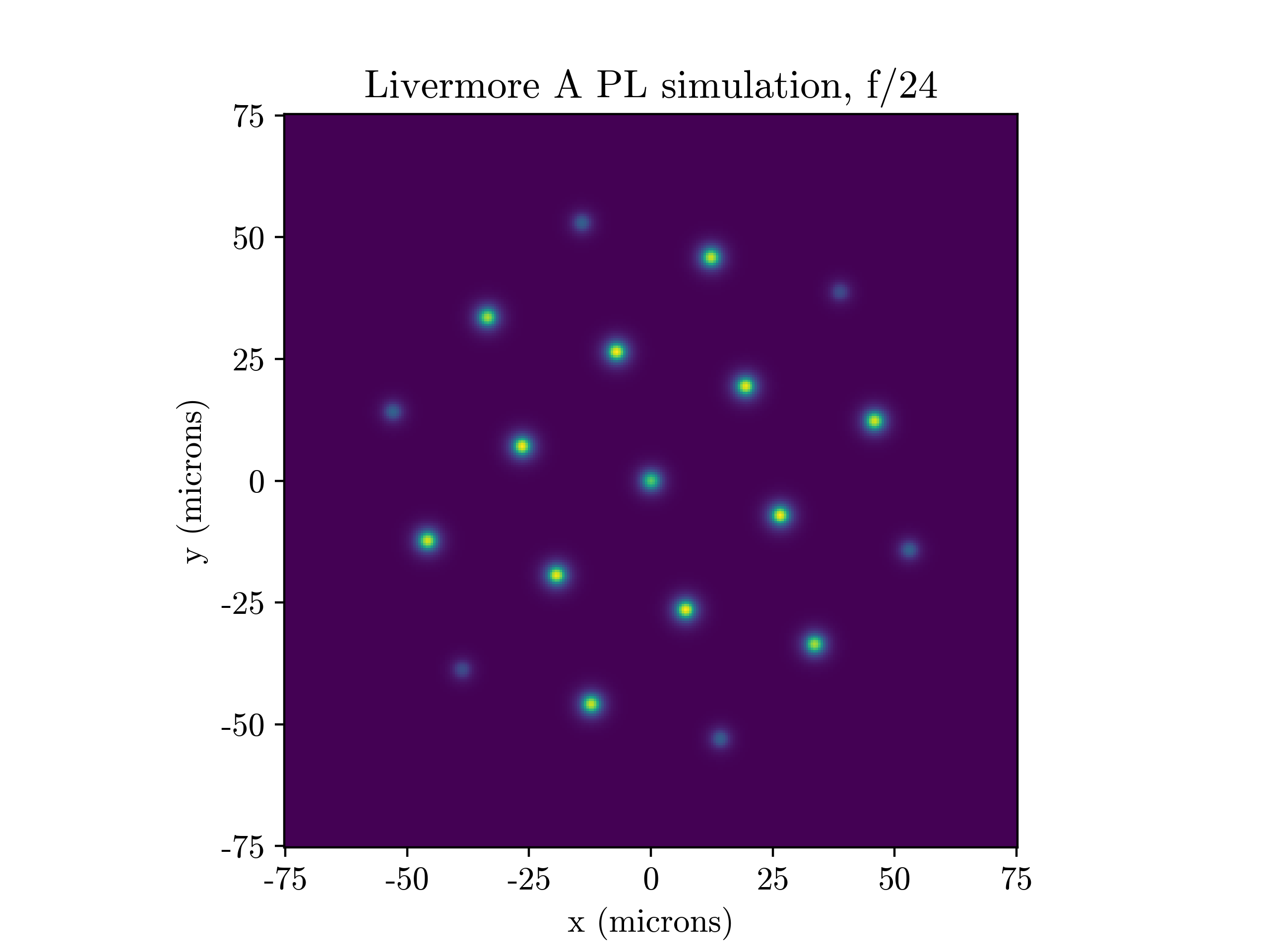}
        \caption{Simulation} \label{fig:simulation}
    \end{subfigure}
    \caption{Qualitative comparison of the LLNL lantern in simulation and the lab. Note that it was not verified that the lab image was obtained with diffraction-limited wavefront quality and thus this lab to simulation comparison should be viewed only as a qualitative one.}
    \label{fig:llnl_imaging}
\end{figure}

We have acquired initial images of the single-mode end (Figure~\ref{fig:real}) and are working on more detailed characterization of the lantern's imaging and wavefront sensing performance. We simulated the design of the standard lantern using the \textit{lightbeam}\cite{lightbeam} Python package and compared this to our lab images. Figure~\ref{fig:simulation} shows the image at a flat wavefront produced by this simulation. We note differences in the relative brightness of the ports that may be caused by imperfections in optical alignment.

\acknowledgments     
A.S. thanks Parth Nobel, William Hartog, and Jonathan Lin for valuable discussions. The document number is LLNL-PROC-2010343. This work was performed under the auspices of the U.S. Department of Energy by Lawrence Livermore National Laboratory under Contract DE-AC52-07NA27344. 

\bibliography{report} 
\bibliographystyle{spiebib} 

\end{document}